\documentclass[aps,pra,nofootinbib,preprint,
showpacs,amsmath,amssymb,floatfix,showpacs]{revtex4}
\usepackage{graphicx}

\begin{document}

\title{Casimir force between a half-space and a plate of finite thickness}         

\author{Johan S. H{\o}ye}        

\date{\today}          

\address {Department of Physics, Norwegian University of Science and Technology, 7491 Trondheim, Norway,}
\author{Iver Brevik}
\address{Department of Energy and Process Engineering, Norwegian University of Science and Technology, 7491 Trondheim, Norway}

\begin{abstract}

Zero-frequency Casimir theory is analyzed from different viewpoints, with the aim of obtaining further insight in the delicate Drude-plasma  issue that turns up when one considers  thermal corrections to the Casimir force.  The problem is essentially that the plasma model, physically inferior in comparison to the Drude model since it leaves out dissipation in the material, apparently gives the best results when comparing with recent experiments.  Our geometric setup is quite conventional,  namely a  dielectric plate separated from a dielectric half-space by a vacuum gap,  both   media  being made of the same material. Our investigation is divided into the following categories: (1)  Making  use of the statistical mechanical method developed by H{\o}ye and Brevik  (1998), implying that the quantized electromagnetic field is replaced by interaction between dipole moments oscillating in harmonic potentials,  we first  verify that the Casimir force is in agreement with the Drude prediction.  No use of Fresnel's reflection coefficients is made at this stage. (2) Then turning  to the field theoretical description implying use of  the reflection coefficients,  we derive  results in agreement with the forgoing when first setting the frequency equal to zero, before letting the permittivity  becoming large.  With the plasma relation the reflection coefficient for TE zero frequency modes depend on the component of the wave vector parallel to the surfaces and lies between 0 and 1.   This contradicts basic electrostatic theory. (3) Turning to high permeability magnetic materials the TE zero frequency mode describes the static magnetic field in the same way as the TM zero frequency modes describe the static electric fields in electrostatics. With the plasma model magnetic fields, except for a small part, can not pass through metals. i.e.~metals effectively become superconductors. However, recent experimental results  clearly favor the plasma model. We shortly discuss a possible explanation for this apparent conflict with electrostatics.
\end{abstract}
\maketitle

\section{Introduction}
\label{sec1}

The finite temperature issue related to the Casimir effect has turned out to be surprisingly difficult to clear out \cite{hoye98,hoye03,hoye06,hoye07,brevik08,brevik14,bostrom00,sushkov11,decca05} (this list of reference is not meant to be exhaustive; readers interested in general reviews of the Casimir effect, may consult Refs.~\cite{milton01,bordag09,dalvit13,simpson15}). The problem arises already in the simple standard setup where there are two parallel dielectric or metallic plates separated by a width $a$. We will assume that the two media are of the same material, that the temperature $T$ is the same everywhere, and we will be concerned only with the transverse force (pressure) per unit surface, called $f_s$.

A key element in the theoretical description of the effect is the choice of dispersion relation in the material. One would expect that the Drude relation
\begin{equation}
\varepsilon(i\zeta)=1+\frac{\omega_p^2}{\zeta(\zeta+\nu)} \label{0}
\end{equation}
is the natural choice here, where  $\zeta$  denotes the imaginary frequency,  $\omega_p$  the plasma frequency, and $\nu$ is a  dissipative term that  describes ohmic resistance. An alternative dispersion relation has however often been proposed in the literature, namely the plasma relation which corresponds simply to setting  $\nu=0$ in Eq.~(\ref{0}).

Unless very special effects are at play in this problem, one would think that the Drude alternative is physically the most correct one. And we ought to emphasize that there is not any fundamental conflict with basic laws of thermodynamics here; in particular, there is no conflict with the Nernst theorem. A detailed calculation has shown that the Nernst theorem is satisfied when the temperature approaches zero \cite{hoye07}. This point ought to be mentioned, as statements to the contrary has repeatedly been made in the literature.

The most important point is however: what is the experimental status concerning the Casimir force?  Thanks to theoretical developments of
Bostr{\"o}m and Sernelius \cite{bostrom00} and others it has been recognized that according to the Drude relation there is no contribution to the force from the TE (transverse electric) zero frequency mode. This gives in turn rise to a characteristic dependence of the force with the separation width at room temperature: at large separations the Drude-related prediction is only one half of the plasma-related prediction. Moreover, for metals the force actually decreases with increasing temperature in a certain temperature interval before it again increases to reach the classical limit for $T\rightarrow \infty$ where only the zero frequency mode contributes.

There are thus possibilities to distinguish experimentally between the two models for the dispersion. The measurements of Sushkov {\it et al.} \cite{sushkov11} were made for a large range of distances, from $0.7$ to $7.3~\mu$m, and were found to agree with the Drude prediction to a good accuracy.  This is as we would expect beforehand. The situation is however more complex: the experiment needed large subtractions to be done in view of the so-called patch potentials, and so brought in some uncertainty regarding the conclusion \cite{klimchitskaya12}. Moreover there have been other experiments, such as the accurate ones of Decca {\it et al.} \cite{decca05}, which have obtained results agreeing with the plasma model instead. This disagreement between experimental groups has left the community in a state of uncertainty for some years.

A new and interesting  approach to this problem is to assume  a thin metal plate of varying thickness, typically Au (gold), overlaying a  thicker plate of a different material like the transition metal Ni (nickel) which is ferromagnetic. Then the small difference in Casimir force can be measured as a variation in it when the plate moves near a metal sphere that can be either Ni or non-magnetic Au. This idea was proposed by Bimonte \cite{bimonte14,bimonte14a,bimonte15a}, and is generalized to the case of unequal temperatures for the two plates also \cite{bimonte15}. (The same kind of technique has been made use of  in the search for non-Newtonian gravity in the submicron range \cite{decca05a,chen14}.) The argument goes roughly as follows: the important point is whether the TE $\omega=0$ mode contributes to the force or not. The TM (transverse magnetic) $\omega=0$ mode, where $\omega$ is frequency, represents static electric fields; they are  screened out and do not contribute to the force. Likewise, the TE zero frequency mode represents static magnetic fields, and can "see" through the thin gold layer if the penetration length  is larger than the width $w$ of the layer. In the Drude approach, this mode is in principle detectable. In the plasma approach, however, this mode only gives a very small contribution,  since  when modeled as a non-dissipative plasma, gold effectively behaves as a superconductor and screens out the static magnetic fields. This method is also proposed for non-magnetic materials like Si (silicon). There the expected difference in Casimir force for the plasma snd Drude models are smaller. With this method, the Fresnel reflection coefficients are used at the plane boundaries between the media.

Conventionally, the Casimir force between two half-spaces is found using the  Lifshitz formula in combination with the  reflection coefficients of the two media. One integrates over transverse wave vectors (transverse to the normal of the plates) and sums over Matsubara frequencies (imaginary frequencies). For {\it layered} media it might at first sight seem reasonable to generalize to this situation by merely inserting the corresponding reflection coefficients. This was done in Refs.~\cite{bimonte14,bimonte14a,bimonte15a,bimonte15} to draw conclusions about interpretation of experimental results. Earlier Pirozhenko and Lambrecht used the same method to evaluate the reduction of Casimir force in a variety of situations with Si and VO$_{\rm 2}$ plates (or films) of finite thickness \cite{porozhenko08}. The Si plates had varying degrees of doping besides its intrinsic case. The VO$_{\rm 2}$ was considered both above and below the critical temperature that separates its metallic and a non-metallic version. A central issue in this connection, as mentioned,  is the possibility to distinguish between the Drude and plasma models for a metal.

By such experiments with ferromagnetic Ni only a small variation in the force has been measured in agreement with results evaluated using the plasma model \cite{bimonte15b}. For non-magnetic materials similar experimental results are not available so far, to our knowledge. Anyway, this has resulted in the firm conclusion that the plasma model, not the Drude one, is the correct model to evaluate the Casimir force for metals.

 A problem with this conclusion is that real metals have after all a finite conductivity, which is neglected in the plasma model. Accordingly, with the plasma model real dielectric data should be irrelevant when obtaining  the proper Casimir force. This seems physically questionable.  It may seem natural therefore,  to reconsider the  straightforward extension of the Lifshitz formula to layered structures. We will only consider the $\omega=0$ limit for simplicity.

 Another point worth attention is the role of the electric and magnetic {\it energies} when $\omega \rightarrow 0$. Consider for definiteness a plane wave falling normally upon a metallic surface. The penetration depth is $\delta=(\mu_0\omega \sigma/2)^{-1/2}$ where $\sigma$ is the conductivity, thus $\delta$ is large in the low frequency limit. In the interior of the metal  the electric field is $E \sim (\omega \delta/c)e^{-x/\delta} \rightarrow 0$ when $\omega \rightarrow 0$. The electric field energy thus goes to zero, and so cannot play any role for the experiment. The magnetic field energy exceeds the electric field energy by a factor $\sigma/(\varepsilon \omega) \gg 1$, and may from this argument be of importance.

 In our calculation below we will first make use of the  the statistical mechanical method used by H{\o}ye and Brevik to  rederive the appropriate version of  Lifshitz formula for the present configuration for non-magnetic materials \cite{hoye98}.  The basis for this method is the induced interaction between a pair of polarizable particles. Then the quantized electromagnetic field is replaced by the dipolar interaction between dipole moments that oscillate in harmonic potentials. In view of the path integral formalism for quantized systems this may also cover the general situation with the time-dependent radiating dipole-dipole interaction where Fourier transform in imaginary time is utilized \cite{feynman53}. But this more demanding situation will not be considered as here we are mostly interested in the electrostatic limit. The pair of polarizable particles is then extended to a pair of polarizable media, e.g. plates. At low densities the resulting force must again be the sum of contributions from pairs of particles. But for higher densities, interactions between dipole moments must be taken into account. Thus one will need the resulting pair correlation function between  particles, one particle in each medium. As shown in Ref.~\cite{hoye98} this, apart from a simple factor, is the Green function of the macroscopic electromagnetic problem. For simplicity we will limit ourselves to the static case in the following, as mentioned.
Our method in Sec.~\ref{sec3} does not involve use of the Fresnel reflection coefficients.

For the electrostatic $\omega=0$ situation we recover the known result by use of reflection coefficients; there is  is no TE zero mode for a non-magnetic system. We will in Sec.~\ref{sec5} discuss in more detail how the plasma model is in conflict with the absence of static electric fields in metals.

Apparently, for some reason, the relatively large difference for magnetic systems seems to favor the plasma description in conflict with the electrostatic picture. A possible reason for this may be the absence of thermal equilibrium in the experimental setup. We discuss this possibility shortly, in Sec~\ref{sec6} dealing with magnetic materials. In the final Sec.~\ref{sec7} we summarize our results.

\ \\


\section{Basic formalism}
\label{sec2}

Consider a pair of harmonic oscillators with oscillator coordinates $s_1$ and $s_2$ (one dimension for simplicity) with potential energy $(s_1^2+s_2^2)/(2\alpha)$ where $\alpha$ is polarizability. They interact via an interaction $\psi s_1 s_2$. The partition function for this system is
\begin{equation}
Z=\int\exp{\left(-\frac{\beta}{2\alpha}(s_1^2+s_2^2)-\beta\psi s_1 s_2\right)}\,ds_1 ds_2=\frac{\pi\alpha}{\beta\sqrt{1-(\alpha\psi)^2}}
\label{1}
\end{equation}
with $\beta=1/(k_B T)$ where $k_B$ is Boltzmann's constant and $T$ is temperature. The total free energy $F_{tot}$ is given by $-\beta F_{tot}=\ln Z$. (The kinetic energy of the classical system has been disregarded here as it does not depend upon $\psi$.) The contribution to the free energy due to the pair interaction $\psi$ is clearly
\begin{equation}
-\beta F=- \frac{1}{2}\ln (1-(\alpha\psi)^2)=\frac{1}{2}\sum\limits_{n=1}^\infty\frac{1}{n}(\alpha\psi)^{2n}.
\label{2}
\end{equation}
The latter series expansion is Eq.~(3.1) of Ref.~\cite{hoye98} where it is explained in terms of the graph structure of statistical mechanics. 
The graphs corresponding to expression (\ref{2}) is the ring graphs occurring in the so-called $\gamma$-ordering for long range forces, $\gamma$ being the inverse range of interaction \cite{hemmer64, lebowitz65,hoye74}, Also it turns out that these graphs yield the exact result for coupled harmonic oscillators \cite{hoye80}.

The Casimir force follows by differentiation of the free energy with respect to the distance $a$ between the two bodies
\begin{equation}
K=-\frac{\partial F}{\partial a}=\frac{1}{\beta}\frac{\alpha\psi\alpha\,\partial\psi/\partial a}{1-(\alpha\psi)^2}.
\label{3}
\end{equation}
In this expression the $\partial\psi/\partial a$ is the derivative of the (dipole) interaction while the remaining part is the pair correlation function for the fluctuating dipole moments of the two particles, one in each medium.

Expressions (\ref{1}) and (\ref{2}) for a pair of particles can formally be extended to a pair of planes. Then the Fourier transform of the dipole-dipole interaction is utilized.
The static dipole-dipole interaction between a pair of dipole moments ${\bf s}_1$ and ${\bf s}_2$ separated by a distance ${\bf r}$ is ($r>0$)
\begin{equation}
\phi(12)=-\frac{s_1 s_2}{r^3}D(12), \quad D(12)=3({\bf\hat r}\cdot{\bf \hat s}_1)({\bf\hat r\cdot\bf \hat s}_2)-{\bf\hat s}_1\cdot{\bf \hat s}_2
\label{11}
\end{equation}
where the hat denotes unit vectors. Its full Fourier transform in 3 dimensions is
\begin{equation}
\tilde\phi(12)=\frac{4\pi}{3}s_1 s_2 \tilde D(12), \quad \tilde D(12)=3({\bf\hat k}\cdot{\bf \hat s}_1)({\bf\hat k\cdot\bf \hat s}_2)-{\bf\hat s}_1\cdot{\bf \hat s}_2.
\label{12}
\end{equation}
 However, with planes normal to the $z$-direction translational symmetry is lost in that direction, so we have to transform back to $z$-space. Thus the dipole-dipole interaction of interest becomes Eq.~(4.3) of Ref.~\cite{hoye98} ($z=z_2-z_1\neq 0 $)
\begin{equation}
\hat\phi(12)=-2\pi s_1 s_2\frac{e^{-q|z|}}{q}({\bf\hat h}\cdot{\bf \hat s}_1)({\bf\hat h\cdot\bf \hat s}_2), \quad q^2=k_\perp^2=k_x^2+k_y,
\label{13}
\end{equation}
\begin{equation}
{\bf h}={\bf h}_\pm=\{ik_x,ik_y,\pm q\}, \quad \mbox{depending on}\quad z\gtrless 0.
\label{14}
\end{equation}

\section{Dielectric half-plane and plane of finite thickness}
\label{sec3}

Now consider a dielectric half-plane for $z<0$ and a corresponding parallel plane for $a<z<b$ surrounded by vacuum. At low particle number density $\rho$ (same density and polarizability in both planes for simplicity) the resulting Casimir force will be the direct sum from all pairs of particles as the denominator in expression (\ref{3}) can be disregarded. The resulting force density $f_s=f_{surf}$ is then given by Eq.~(4.4) in Ref.~\cite{hoye98} except that in the present case one of the integrations is limited to $a<z<b$
\begin{equation}
f_s=\frac{1}{\beta}\frac{1}{(2\pi)^2}\int dk_xdk_y\int\limits_a^b\int\limits_{-\infty}^0 dz_1dz_2(\beta\rho)^2\langle\hat\phi(12)(-q\hat\phi^*(12))\rangle.
\label{15}
\end{equation}
The factor $q$ is from the derivative of the potential according to Eq.~(\ref{3}). For  particles with polarizability $\alpha_0$ Eq.~(4.5) of Ref.~\cite{hoye98} will be as before with thermal averages $\beta\langle s_1^2\rangle=\beta\langle s_2^2\rangle=3\alpha_0$ (for oscillations in the potential $-s^2/(2\alpha_0)$ in 3 dimensions)
\begin{equation}
H=\langle({\bf\hat h\cdot\bf \hat s}_i)({\bf\hat h\cdot\bf \hat s}_i)\rangle=\frac{1}{3}{\bf h\cdot h}^*=\frac{1}{3}(k_\perp^2+q^2)=\frac{2}{3}q^2, \quad (i=1,2).
\label{16}
\end{equation}
With the new limits of integration integral (4.5) of Ref.~\cite{hoye98} now becomes
\begin{equation}
I=\int\limits_a^b\int\limits_{-\infty}^0\left (\frac{e^{-q(z_2-z_1)}}
{q}\right)^2\, dz_1dz_2=\frac{e^{-2qa}}{4q^4}(1-e^{-2q(b-a})
\label{17}
\end{equation}
For a dilute medium $4\pi\rho\alpha_0=\varepsilon-1$ where $\varepsilon$ is permittivity. Altogether, when inserting interaction (\ref{13}) into expression (\ref{15}) where Eqs.~(\ref{16})-(\ref{17}) are further inserted, one finds by use of $\int dk_xdk_y=2\pi\int q\,dq$, that Eq.~(4.9) of \cite{hoye98} is modified into (with $3y=4\pi\rho\alpha_0$)
\begin{equation}
f_s=-\frac{1}{2\pi\beta}\int\limits_0^\infty q^2\,dq\,\left(3\frac{3y}{2}\right)^2IH^2=-\frac{1}{2\pi\beta}\int\limits_0^\infty q^2\,dq\,\left(\frac{\varepsilon-1}{2}\right)^2e^{-2qa}(1-e^{-2q(b-a)}).
\label{19}
\end{equation}

For general permittivity there will be induced fields, and the sought correlation function follows from solution of the corresponding electrostatic problem. Thus the electric field created by a dipole ${\bf s}_1$ (for $\varepsilon=1$) can be written like Eqs.~(5.1) and (5.2) of \cite{hoye98}
\begin{eqnarray}
{\bf E}=Le^{-q|z|}{\bf h}, \quad L=2\pi s_1\frac{1}{q}({\bf h\cdot \hat s_1}).
\label{20}
\end{eqnarray}
\begin{equation}
\label{}
\end{equation}
So for a dipole moment located at $z=z_0<0$ we now have
\begin{displaymath}
{\bf E}e^{-qz_0}/L=\left\{\begin{array}{cccc}
&\frac{1}{\varepsilon}e^{-qz}{\bf h}_+&+Be^{qz}{\bf h}_-, \quad &z_0<z<0\\
&Ce^{-qz}{\bf h}_+&+C_1e^{qz}{\bf h}_-, \quad &0<z<a\\
&De^{-qz}{\bf h}_+&+D_1e^{qz}{\bf h}_-, \quad &a<z<b\\
&Fe^{-qz}{\bf h}_+,& \quad &b<z,
\end{array} \right.
\label{21}
\end{displaymath}
which generalizes Eq.~(5.3) of \cite{hoye98}.

At the three interfaces $z=0,a,b$ the tangential components of ${\bf E}$ and the normal component of ${\bf D}=\varepsilon{\bf E}$ must be continuous. This gives equations
\begin{eqnarray}
\nonumber
\frac{1}{\varepsilon}+B&=&C+C_1\\
1-\varepsilon B&=&C-C_1
\label{22}
\end{eqnarray}
etc. Here we may eliminate $B$ to obtain
\begin{equation}
2=(\varepsilon+1)C+(\varepsilon-1)C_1.
\label{23}
\end{equation}
Likewise for the two other interfaces we find
\begin{eqnarray}
\nonumber
2C&=&(\varepsilon+1)D-(\varepsilon-1)D_1e^{2qa}\\
\nonumber
2C_1&=&-(\varepsilon-1)De^{-2qa}+(\varepsilon+1)D_1\\
\label{24}
2\varepsilon D&=&(\varepsilon+1)F\\
\nonumber
2\varepsilon D_1&=&(\varepsilon-1)Fe^{-2qb}=2\frac{\varepsilon-1}{\varepsilon+1}De^{-2qb}.
\end{eqnarray}
With this the $C$ and $C_1$ can be expressed in terms of $D$ alone by which Eq.~(\ref{23}) gives the following relation for $D$
\begin{equation}
4=(\varepsilon+1)^2\left[1-\left(\frac{\varepsilon-1}{\varepsilon+1}\right)^2(e^{-2qa}-e^{-2qb}+e^{-2q(b-a)})\right]D.
\label{25}
\end{equation}
Of special interest is the case of a metal for which $(\varepsilon-1)/(\varepsilon+1)\rightarrow1$. The the right hand side can be factorized to obtain
\begin{equation}
4=(\varepsilon+1)^2(1-e^{-2qa})(1-e^{-2q(b-a)})D.
\label{26}
\end{equation}

As explained in connection with Eq.~(5.5) of \cite{hoye98} the correlation function of a dielectric system deviates from the Green function of the electromagnetic problem by a factor
\begin{equation}
A=\left(\frac{\varepsilon-1}{3y}\right)^2, \quad 3y=\frac{4\pi}{3}\rho\beta\langle s^2\rangle.
\label{27}
\end{equation}
With this the force density $f_s$ is the low density result (\ref{19}) multiplied with $AD$ where the $\varepsilon-1$ term of (\ref{19}) is replaced with its low density value $\varepsilon-1=3y$. We find
\begin{equation}
f_s=-\frac{1}{2\pi\beta}\int\limits_0^\infty q^2\,dq \frac{\Delta^2 e^{-2qa}(1-e^{-2q(b-a)})}{1-\Delta^2(e^{-2qa}-e^{-2qb}+e^{-2q(b-a)})},
\label{27a}
\end{equation}
\begin{equation}
\Delta^2=\left(\frac{\varepsilon-1}{\varepsilon+1}\right)^2.
\label{27b}
\end{equation}
With two half-planes $b\rightarrow \infty$
we thus have the well known Lifshitz formula
\begin{equation}
f_s=-\frac{1}{2\pi\beta}\int\limits_0^\infty q^2\,dq \frac{\Delta^2 e^{-2qa}}{1-\Delta^2 e^{-2qa}},
\label{27c}
\end{equation}
The found expression (\ref{27a}) may also be written in the same form
\begin{equation}
f_s=-\frac{1}{2\pi\beta}\int\limits_0^\infty q^2\,dq \frac{\Delta'^2 e^{-2qa}}{1-\Delta'^2 e^{-2qa}},
\label{27d}
\end{equation}
where with $w=b-a$
\begin{equation}
\Delta'^2=\frac{\Delta^2(1-e^{-2qw})}{1-\Delta^2 e^{-2qw}}.
\label{27e}
\end{equation}

Of special interest is the case of a metal for which $A_0=(\varepsilon-1)/(\varepsilon+1)\rightarrow1$. Then $\Delta'=\Delta=1$ or the right hand side of Eq.~(\ref{25}) can be factorized to obtain
\begin{equation}
4=(\varepsilon+1)^2(1-e^{-2qa})(1-e^{-2q(b-a)})D.
\label{27f}
\end{equation}
With this the Casimir force density (\ref{27}) with a metal plate becomes
\begin{equation}
f_s=-\frac{1}{2\pi\beta}\int\limits_0^\infty q^2\,dq\,\frac{e^{-2qa}}{1-e^{-2qa}}
\label{28}
\end{equation}
as the factor $1-e^{-2q(b-a)}$ cancels. The $D_1$ term may also contribute, but in the electrostatic case it does not since ${\bf h}_+\cdot{\bf h}_-=k_x^2+k_y^2-q^2=0$. 	Thus the result is independent of the thickness $b-a$ of the metal layer and is accordingly the same as  for two metal half-planes. This is then also fully consistent with the electrostatics of continuous media. Electric fields do not penetrate metals.

Taking into account the relation
\begin{equation}
\int_0^\infty \frac{z^{x-1}dz}{e^z-1}=\Gamma(x)\zeta(x), \quad (x>1),
\end{equation}
with $\zeta(x)$ the Riemann zeta-function, we can express $f_s$ as
\begin{equation}
f_s=-\frac{\zeta(3)}{8\pi \beta a^3}. \label{trykk}
\end{equation}

\section{Approach in terms of reflection coefficients}
\label{sec4}

As a reassurance it is of interest to consider the same geometric setup if one uses instead the conventional field theoretical formalism and the reflection coefficients at the boundaries. We again limit ourselves to the static case. Assume that a wave falls from vacuum (subscript zero) towards a dielectric nonmagnetic medium with real permittivity $\varepsilon$. As before, we write the constitutive relations as ${\bf D}=\varepsilon_0\varepsilon {\bf E}, ~{\bf B}=\mu_0{\bf H}$, where $\varepsilon$ is real and constant. The TE and TM reflection coefficients, here called $\Delta_{\rm TE}$ and $\Delta_{\rm TM}$, are
\begin{equation}
\Delta_{\rm TE}=\frac{\kappa -\kappa_0}{\kappa +\kappa_0}, \label{42}
\end{equation}
\begin{equation}
\Delta_{\rm TM}=\frac{\kappa-\varepsilon \kappa_0}{\kappa+\varepsilon \kappa_0}, \label{43}
\end{equation}
where
\begin{equation}
\kappa=\sqrt{k_\perp^2-\varepsilon \omega^2/c^2}, \quad \kappa_0=\sqrt{k_\perp^2-\omega^2/c^2}.
\label{44}
\end{equation}
The Casimir force $f_s$ on the slab of width $w=b-a$ situated at a distance $a$ from the half-space $z<0$ with the same permittivity $\varepsilon$ can conveniently be found, for instance, by simplifying the more general formalism of Ref.~\cite{ellingsen07} pertaining to a dielectric slab situated in a cavity (cf. also the related Ref.~\cite{ellingsen07a} considering multilayered systems in general). The simplification consists in letting the distance to one of the walls go to infinity. We can then write
\begin{equation}
f_s=-\frac{1}{\pi \beta}{\sum_{m=0}^\infty}^\prime \int_0^\infty dk_\perp k_\perp \kappa_0 \left(\frac{1}{d_{\rm TE}}+\frac{1}{d_{\rm TM}}\right),
\end{equation}
where the prime means that the mode $m=0$ is to be taken with half weight, and $d_{\rm TE}$ and $d_{\rm TM}$ are complicated functions which in the present case simplify considerably. We obtain
\begin{equation}
f_s=-\frac{1}{\pi \beta}{\sum_{m=0}^\infty}^\prime \int_0^\infty dk_\perp k_\perp \kappa_0 \left[ \frac{e^{-2\kappa_0 a}}{(\Delta'_{\rm TE})^{-2}-e^{-2\kappa_0a}}+  \frac{e^{-2\kappa_0 a}}{(\Delta'_{\rm TM})^{-2}-e^{-2\kappa_0a}} \right], \label{46}
\end{equation}
where $\Delta'_q$ is a combined reflection coefficient,
\begin{equation}
(\Delta_q')^{-2}=\frac{\Delta_q^{-2}-e^{-2\kappa w}}{1-e^{-2\kappa w}}, \quad q=(\rm TE,TM). \label{47}
\end{equation}
Recall that $\varepsilon$ is assumed real and finite; otherwise arbitrary.

Consider now the static mode, $\omega=0$. From Eqs.~(\ref{42}) and (\ref{43}) it follows that for $k_\perp \neq 0$,
\begin{equation}
\Delta_{\rm TE}=0, \quad \Delta_{\rm TM}=\frac{1-\varepsilon}{1+\varepsilon}.
\end{equation}
Thus the TE mode does not contribute, and we obtain
\begin{equation}
f_s(\omega=0)=-\frac{1}{2\pi \beta}\int_0^\infty k_\perp^2dk_\perp \frac{e^{-2k_\perp a}}{(\Delta'_{\rm TM})^{-2}-e^{-2k_\perp a}},\label{49}
\end{equation}
where
\begin{equation}
(\Delta'_{\rm TM})^{-2}=\frac{\Delta_{TM}^{-2}-e^{-2k_\perp w}}{1-e^{-2k_\perp w}}. \label{50}
\end{equation}
One sees that for $\omega=0$ this is the same as expression (\ref{27e}) obtained with the statistical mechanical method for the electrostatic case. Note that the absence of the TE mode for $\omega=0$ is fully consistent with the stastistical mechanical derivation.

For large values of $\varepsilon$, which is the case of main importance here, there is thus a weak dependence on the width $w$ in the expression (\ref{49}) for the force. But in the limit $\varepsilon \rightarrow \infty$ corresponding to a metal, it is seen that
\begin{equation}
(\Delta'_{\rm TM})^{-2} \rightarrow 1
\end{equation}
smoothly, and the dependence on $w$ goes away. We end up with
\begin{equation}
f_s(\omega=0, \varepsilon \rightarrow \infty)=-\frac{1}{2\pi \beta}\int_0^\infty k_\perp^2dk_\perp \frac{e^{-2k_\perp a}}{1-e^{-2k_\perp a}}=-\frac{\zeta(3)}{8\pi \beta a^3}. \label{52}
\end{equation}
This is precisely the same as our previous expression (\ref{trykk}), now derived in an alternative manner. There is no trace of the mentioned penetration depth $\delta$ in this expression. However, if one first lets $\varepsilon\rightarrow\infty$ or use the Drude relation (\ref{0}) with $\nu=0$ (plasma model) and then let $\omega\rightarrow 0$, then the plasma model non-zero $\Delta_{TE}$ for the TE zero mode appears. This is in conflict with the electrostatic derivation of Sec.~\ref{sec3}. 

\section{Plasma model for metals}
\label{sec5}

The consequence of the plasma model for metals will be the presence of a TE zero  mode. The Casimir force from this mode will be like integral (\ref{27c}) with reflection coefficient given by Eqs.~(\ref{42}) and (\ref{44}). For the Drude model the dielectric constant constant is given by Eq.~(\ref{0}).
When inserted in expression (\ref{44}) for $\zeta=0$ this gives the reflection coefficient $\Delta_{\rm TE}=0$ for the Drude model with $\nu>0$. However, for the plasma model one puts $\nu=0$ by which one for $\zeta\rightarrow 0$ thus finds expression (\ref{42}) with
\begin{equation}
\kappa=\sqrt{k_\perp^2+\omega_p^2/c^2}, \quad \kappa_0=k_\perp,\quad (\omega=0).
\label{62}
\end{equation}
Thus one finds a non-zero reflection coefficient that varies with transverse wave vector ${\bf k}_\perp$. For $k_\perp=0$ one has $\Delta_{\rm TE}=1$ like the electrostatic $\Delta_{\rm TM}=1$ for metals. However, for increasing values of $k_\perp$ ($>0$) the plasma model $\Delta_{\rm TE}$ will start to decrease to become less than one, i.e.~$0<\Delta_{\rm TE}<1$. This has the additional consequence that the metal becomes partially transparent to the TE field at zero frequency. In our opinion this is in conflict with known electrostatics of metals where static electric field should be zero inside in accordance with our derivations in Sec.~\ref{sec3}.

Further, it has been shown that expression (\ref{trykk}) is also the same as the classical result for two half-planes where the metals are replaced with ionic plasma or ionic fluids for which classical statistical mechanics is applied \cite{jancovici04}. With non-zero TE mode this mode would add to expression (\ref{trykk}) by which the resulting Casimir force would be different from the classical result with ionic fluids. This would give a mismatch with a different high temperature classical limit, and it reflects another disturbing feature of the plasma  model. When the limit $\nu\rightarrow 0$ is considered, the resulting force changes discontinuously when $\nu$ reaches 0. To us such a discontinuity seems unphysical. This was also noted in Ref.~\cite{porozhenko08}. 

In Ref.~\cite{bimonte15a} an experiment is proposed that should be able to distinguish between the plasma and Drude models. An alternating layer of Au (gold) and high resistivity Si is covered by a thin layer of conducting Si. Then the alternating Casimir force between this and a Au sphere should be measured. Non-zero Matsubara frequencies will to some extent partially penetrate the metal like conducting Si top layer. However, the crucial difference is that in addition the plasma model TE $\omega=0$ mode will partially penetrate it too while it will be absent for the Drude model. This gives two clearly different results for the alternating Casimir force that are expected to be tested experimentally. Note that the materials considered here are nonmagnetic.

\section{Magnetic materials}
\label{sec6}

In Ref.~\cite{bimonte15a} an experiment with a magnetic material is also proposed to distinguish between the plasma and Drude models. Then an alternative layer of Au (gold) and magnetic Ni is covered by a thin layer of Au. The alternating Casimir force between this and a Ni sphere should be measured. With this setup there should be a magnetic contribution to the Casimir force. Due to slow motion (large damping) of magnetic moments only the lowest $\zeta=0$ Matsubara frequency will contribute. Then with the plasma model one again neglects or puts $\nu=0$ in the dielectric constant, Eq.~(\ref{0}). With this it is found that the Au overlayer cannot be penetrated by the magnetic field to reach the Ni below for $k_\perp=0$. For $k_\perp>0$ a small part goes through like the electric field for this model. The static magnetic field is described by the TE zero mode in the same way as the TM zero mode describes electrostatics. Altogether with the plasma model the Au overlayer effectively acts as a superconductor that prevents magnetic fields to penetrate it. On the contrary with the Drude model ($\nu>0$) the magnetostatic field passes unrestricted through the Au layer.

Experiments have already been performed to measure the influence upon the Casimir force \cite{bimonte15b}. The results of these experiments clearly agrees very well with the evaluations based upon the plasma model. The much larger magnetic force from the Drude model is not seen. To check the influence from the magnetic force the Ni sphere was also replaced by a non-magnetic Au sphere in some of the experiments (with unchanged alternating layer). The difference in results were within experimental accuracy in agreement with the calculated magnetic force. However, with the Au sphere Drude model calculations were also done. But for this situation the difference between the two models was to small to draw conclusions from the experiment.

From the magnetic Ni experiment the firm conclusion was drawn that the plasma model is the correct model. In view of the problems this raises in the electrostatic situation discussed in Sec.~\ref{sec5}, we find that this may be investigated further. By looking more closely into the magnetic case we note some problems also indicated in Refs.~\cite{bimonte15a} and \cite{bimonte15b}. Clearly Ni is ferromagnetic, and in the ferromagnetic state magnetization is present in domains. Ideally they should cancel each other to give zero net magnetization. However, one ends up with an athermal magnetic microstructure with remnant magnetization. Also domains are difficult to change by thermal fluctuations. Thus by the experiment it was necessary to average over magnetic configurations below the Ni sphere in a suitable way to avoid periodic repetitions of the magnetic signature. This indicates that there may be problems with thermal equilibrium as the metal plate below the Ni sphere is moving (rapidly on the nm scale). To obtain thermal equilibrium it is at least necessary for the magnetic field to first penetrate the Au overlayer where currents are induced to counteract the field. The decay of these currents are determined by the $\nu$ which is proportional to resistivity. Although Au like other metals are not superconductors at room temperature, they may act like that for short periods. Thus in this respect the $\nu$ may be neglected such that the plasma model effectively describes the induced magnetic Casimir interaction. However, since these considerations are speculations so far, further investigations are needed.

\section{Conclusion, and final remarks} \label{sec7}

 Let us first summarize the outcomes of the above analysis. Our geometric setup was a dielectric half-plane for $z<0$ and a parallel plate in the region $a<z<b$, made of the same material. On both sides of the plate, $0<z<a$ and $z>b$, we assumed a vacuum. We moreover assumed   finite temperature $T$ throughout, and focused attention on the zero-frequency case, i.e. the static limit.

Our first approach, Secs.~\ref{sec2} and \ref{sec3}, was based on quantum statistical physics, starting out from the partition function $Z$. We followed the method developed in Ref.~\cite{hoye98}. It is worth noticing that the correlation function of a dielectric system deviates from the electromagnetic Green function by a factor $A$, given by Eq.~(\ref{27}). In the metallic limit $\varepsilon \rightarrow \infty$ we derived in this way the expressions (\ref{28}) and (\ref{trykk}) for the force. These expressions are proportional to $T$, inversely proportional to $a^3$, and in view of the metallic limit independent of the plate width $w=b-a$. Further there is no TE zero mode by this derivation. We emphasize that with this statistical mechanical procedure we did not make use of the Fresnel reflection coefficients at all.

In our second approach of Sec. \ref{sec4} we made, by contrast, explicit use of the reflection coefficients.  The full surface force is given by Eqs.~(\ref{46}) and (\ref{47}), and the $\omega=0$ contribution is given by Eqs.~(\ref{49}) and (\ref{50}) in agreement with the results of Sec.~\ref{sec3}. Thus in the metallic limit $\varepsilon \rightarrow \infty$ we also recovered the same expression (\ref{52}) as before for the Casimir force per unit area.

Note the way in which the limiting procedure in Sec.~\ref{sec4} was applied: we {\it first} put $\omega=0$, {\it thereafter} took the limit of high permittivity. This is precisely the characteristic property for how to calculate the {\it Drude} expression for the force. We can thus make the following important conclusion: the Drude theory gives results in agreement with the statistical mechanical approach.

As noted, the zero frequency TE mode does not contribute to the force in the Drude theory. For non-magnetic systems only the electrostatic dipole-dipole interaction is present, and thus no TE component appears. If we were to adopt the plasma theory instead, the zero frequency TE mode would contribute, equally to the corresponding TM mode, and the result (\ref{52}) for the force would have to be multiplied with a factor 2 (for ideal metals for which  $\Delta_{\rm TE}=1$ for all $k_\perp$, i.e.~$\varepsilon(\omega)=\infty$).

In Sec.~\ref{sec5} problems with the plasma model are further discussed. With this model a part of electrostatic fields can penetrate metals. This disagrees with the knowledge that electric fields can not enter metals. Further the plasma model disagrees with the classical high temperature limit for the Casimir force between ionic plasma. Anyway, experiments as described may make this more clear.

In Sec.~\ref{sec6} the results of experiments with ferromagnetic Ni discussed. They clearly favor the plasma model in conflict with the discussion of Sec.~\ref{sec5}. However, we note that Ni is ferromagnetic, and the system is athermal with magnetic domains. Thus thermal equilibrium is not reached due to these domains and/or the rapid relative motion on the nm-scale. In this respect the Au layer effectively acts as a superconductor such that the damping coefficient $\nu$ can be neglected. But the latter arguments are speculations. So further investigations are needed to clear up the matter, especially experimental results on non-magnetic metals.

\end{document}